\documentclass[12pt]{iopart}
\usepackage{graphicx}
\usepackage{cite}
\usepackage{amsfonts}

\begin{document}

\title[Absorbing boundaries in the conserved Manna model]
      {Absorbing boundaries in the conserved Manna model}
\author{Arthur Hipke, Sven L{\"u}beck, and Haye Hinrichsen}
\address{Universit\"at W\"urzburg\\
	 Fakult\"at f\"ur Physik und Astronomie\\
         D-97074 W\"urzburg, Germany}

\def\d{\mathrm{d}}
\def\xvec{{\bf x}}

\begin{abstract}
The conserved Manna model with a planar absorbing boundary is studied in various space dimensions. We present a heuristic argument that allows one to compute the surface critical exponent in one dimension analytically. Moreover, we discuss the mean field limit that is expected to be valid in $d>4$ space dimensions and demonstrate how the corresponding partial differential equations can be solved.
\end{abstract}

\submitto{Journal of Statistical Mechanics: Theory and Experiment}
\pacs{05.50.+q, 05.70.Ln, 64.60.Ht}


\parskip 2mm 

\def\d{{\rm d}}
\def\te{{t_{\scriptscriptstyle \hspace{-0.3mm} e}}}
\def\tf{{t_{\scriptscriptstyle \hspace{-0.3mm} f}}}
\def\Ps{{P_{\scriptscriptstyle \hspace{-0.3mm} s}}}
\def\Pe{{P_{\scriptscriptstyle \hspace{-0.3mm} e}}}

\section{Introduction}

Absorbing phase transitions are a particular class of nonequilibrium phase transitions (see \cite{MarroDickman99,Hinrichsen00,Odor04,Lubeck04} for recent reviews). These transitions continue to attract attention since they have no equilibrium counterparts and occur even in one-dimensional systems. The stochastic process of directed percolation (DP) is recognized as a paradigmatic example of absorbing phase transitions. According to the well-known conjecture by Janssen and Grassberger\cite{Janssen81a,Grassberger82} a continuous phase transition should belong to the universality class of DP if the corresponding model has a scalar order parameter with a single absorbing state, evolves by short-range interactions without quenched disorder and has no unconventional symmetries and conservation laws. Non-DP behavior is expected to occur if one of these conditions is violated. In particular, particle conservation leads to an autonomous universality class of nonequilibrium phase transitions which is usually referred to as the Manna universality class. In the recent years, the scaling properties of the Manna universality class were investigated in several mostly numerical works~(see \cite{Lubeck04} and references therein). The same type of critical behavior occurs in various models such as the Manna sandpile model~\cite{Manna91a} (which was introduced in the context of self-organized criticality), the conserved lattice gas (CLG)~\cite{RossiEtAl00a} and the conserved threshold transfer process (CTTP)~\cite{MendesEtAl94}. In addition, the Maslov-Zhang sandpile~\cite{MaslovZhang96} and the Mohanty-Dhar sandpile~\cite{MohantyDhar02}, which were believed to belong to DP, were shown to be a member of the Manna universality class as  well~\cite{BonachelaEtAl06,MohantyDhar07,BonachelaMunoz08}. Moreover, the Manna universality class is of fundamental interest since it connects the critical behavior of absorbing phase transitions with the critical state of self-organized critical (SOC) systems~\cite{Bak87a}. Actually, SOC sandpile models can be considered as driven-dissipative versions of (closed) systems exhibiting absorbing phase transitions~\cite{Bak87a}. An intriguing consideration of the interplay between absorbing phase transitions and SOC in case of the CTTP can be found in~\cite{Lubeck04}. 

As in equilibrium statistical mechanics, critical systems out of equilibrium respond strongly to the presence of boundaries. For systems in the DP class, the problem of  boundary effects was investigated thoroughly in a series of papers, where numerical simulations, field theoretical  approaches~\cite{JanssenEtAl88,LauritsenEtAl98,FroedhEtAl98a,FroedhEtAl01a}, series expansions techniques~\cite{EssamEtAl96,Jensen99}, as well as density matrix renormalization group methods~\cite{CarlonEtAl99} have been applied. These studies focused on two different types of boundary conditions, namely, so-called \textit{absorbing walls}, where activity is set to zero by removing particles, and \textit{active walls}, where activity is artificially held at a high level. Usually absorbing walls generate a non-trivial surface-critical behavior which can be described in terms of an additional surface critical exponent $\beta_s$. On the other hand, the impact of active walls can be explained in terms of the standard critical exponents only. In all cases the bulk critical behavior remains unaffected, i.\,e.~the bulk scaling behavior still belongs to the ordinary DP universality class. Analogous results were obtained for the so-called parity-conserving universality class of absorbing phase transitions~\cite{FroedhEtAl01a}.

Such boundary effects may also be studied within the Manna universality class. Compared to DP, 
the implementation of an absorbing wall in the Manna class is more subtle since the dynamics at the boundary has to preserve the global number of particles. Furthermore, the existing knowledge about boundary effects in the Manna class is still limited and systematic studies started only recently~\cite{BonachelaMunoz07}. In the present paper we extend these studies, discussing the scaling behavior near the boundary, computing the surface exponent exactly in $d=1$, and analyzing the mean-field limit which is expected to describe the system above the upper critical dimension $d_c=4$.

\section{Numerical results}

The Manna model is defined on a lattice with $L^d$ sites,
where each site $i$ is associated with an integer number $n_i$ 
representing the local number of particles. 
Sites with $n_i<N_c$ ($n_i \geq N_c$) grains are considered as inactive (active), 
where $N_c$ is a stability threshold, usually $N_c=2$. 
The Manna model evolves by parallel updates by simultaneously redistributing 
all particles at active sites to randomly chosen nearest neighbor sites, 
which in turn may become active.
Note that this update procedure conserves the total number of particles~$N$
if periodic boundary conditions are applied.
Each parallel update counts as a single time step.
The Manna model is known to display a continuous phase transition from 
a fluctuating active phase into frozen configurations (lacking of active sites),
where the particle density $\phi=N/L^d$ plays the role of a control parameter. 
The phase transition takes place at the critical threshold
$\phi_c$ which depends on the spatial dimension, 
the lattice structure, as well as on the particular value 
of $N_c$~\cite{Lubeck04}.
In order to preserve the number of particles, the absorbing wall is implemented 
by removing all particles from \textit{active} boundary sites and adding them 
to randomly chosen sites in the bulk of the system.

\begin{figure}[t]
\includegraphics[width=75mm]{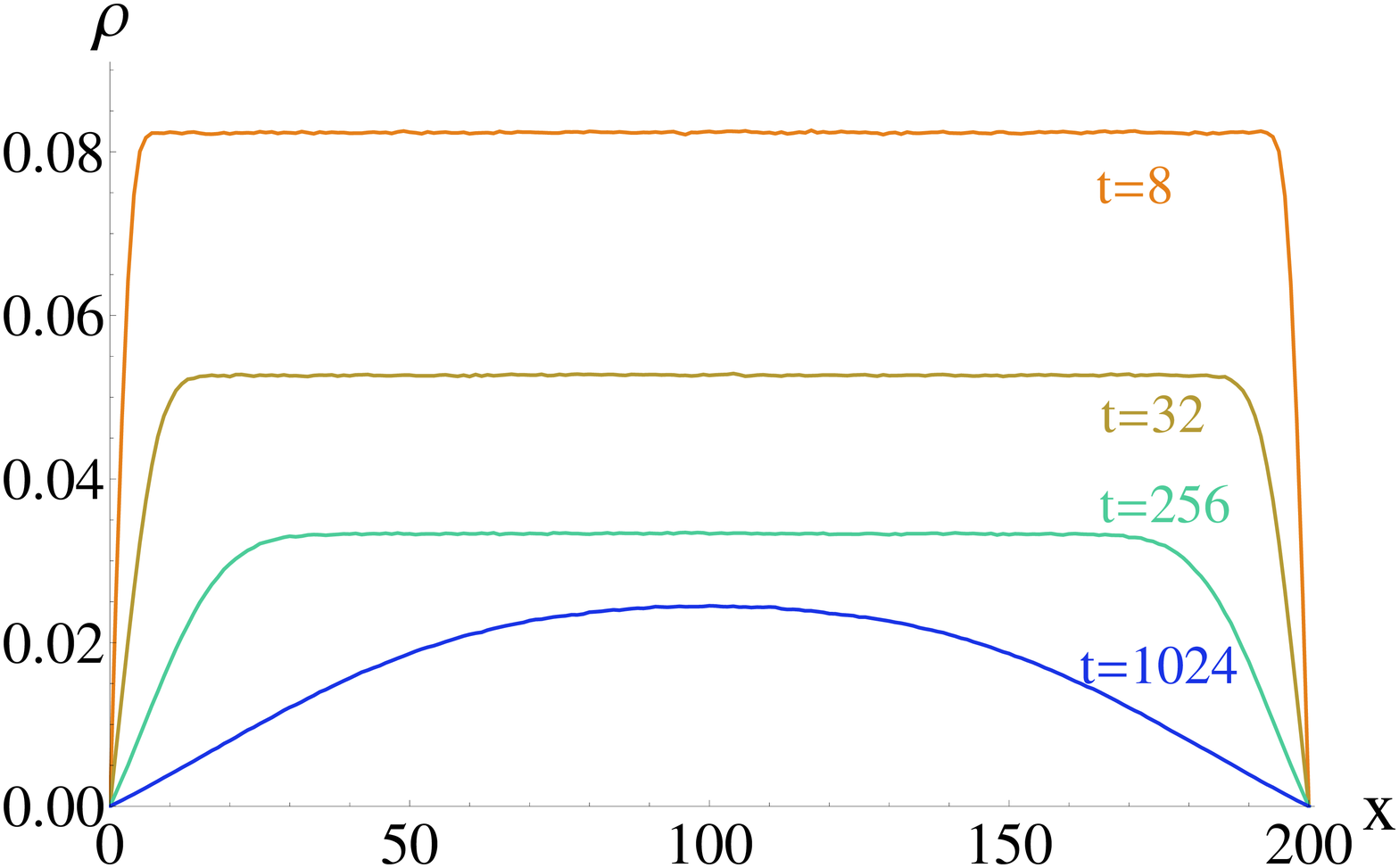}
\hglue 5mm
\raggedright
\includegraphics[width=75mm]{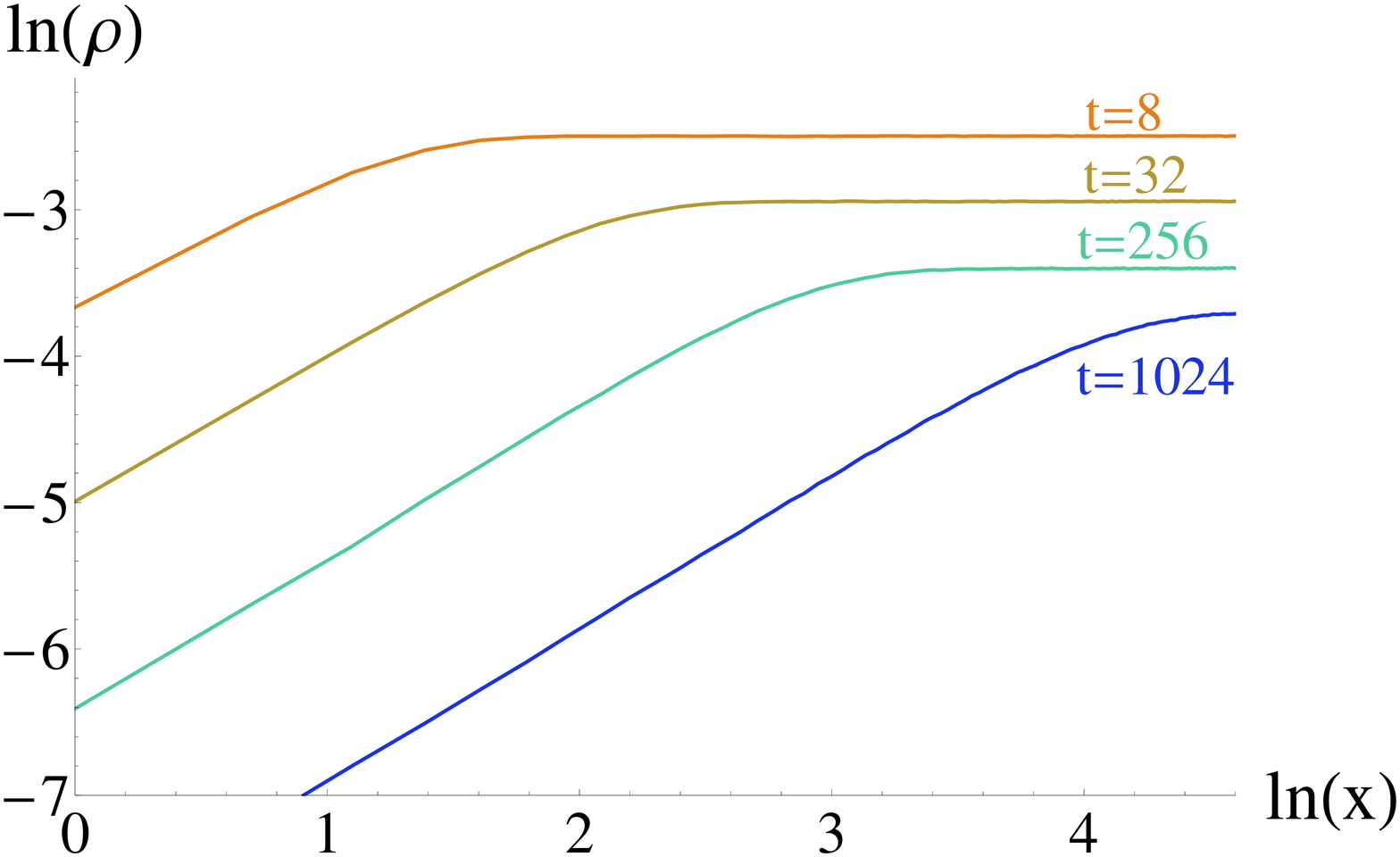}
\caption[Conserved Manna model with an absorbing boundary]{Conserved Manna model in two dimensions with an absorbing boundary in form of a line: Influence of an absorbing boundary on the density of active sites in a linear (left) and a logarithmic (right) plot. Parameters: Lattice with $N = 200 \times 200$ and periodic boundary conditions simulated at $\phi = \phi_c$ and averaged over $65536$ runs. The corresponding results in 1 and 3 dimensions are qualitatively similar.}
\label{fig:2dProfil}
\end{figure}

First we studied the spatio-temporal impact of absorbing boundaries 
in different space dimensions for systems starting with a homogeneously 
active state. 
Fig.~\ref{fig:2dProfil} shows the decay of the density of active 
sites as a function of the perpendicular distance to the wall at 
different times. 
As expected, the absorbing wall creates a depletion zone that grows 
with time as $t^{1/z}$. 
The density of active sites inside the depletion zone scales as
\begin{equation}
\label{ScalingLimit}
\rho \sim x^{{\kappa_s}}\qquad \mbox{ for }\,\, x \ll t^{1/z}, 
\label{eq:kappa}
\end{equation}
where $x$ denotes the distance to the boundary and $\kappa_s$ is a new exponent.
According to the standard scaling theory for surface-critical 
phenomena this suggest the scaling form
\begin{equation}
\label{ScalingForm}
\rho(x,t) \;=\; t^{-\alpha} \,R(x/t^{1/z})\,,
\end{equation}
where $R$ is a scaling function with the asymptotic behavior 
\begin{equation}
\label{AsymptoticBehavior}
R(\xi)\;\sim\;\left\{
\begin{array}{ll}
\xi^{\kappa_s} & \mbox{ for } \xi\ll 1 \\
const & \mbox{ for } \xi\gg 1
\end{array}
\right.\,.
\end{equation}
Our numerical simulations lead to the estimates
\begin{equation}
{\kappa_s}\;=\; \left\{
\begin{array}{ll}
1.00(2) & \mbox{ for } d=1 \\
1.10(2) & \mbox{ for } d=2 \\
1.40(15) & \mbox{ for } d=3 \\
\end{array}
\right.\,.
\end{equation}
The asymptotic behavior of $R$ implies that the activity $\rho_s$ next 
to the boundary, e.g. at a site adjacent to the wall, decays as 
\begin{equation}
\rho_s(t) \sim t^{-\alpha_s}\,.
\end{equation}
Using the scaling form~(\ref{ScalingForm}) with the asymptotic behavior~(\ref{AsymptoticBehavior}) for $\xi\ll 1$ one is led to
\begin{equation}
\rho_s(t) = \rho(1,t) = t^{-\alpha} \Bigl( \frac{1}{t^{1/z}} \Bigr)^{\kappa_s} = t^{-\alpha-\kappa_s/z}
\end{equation}
which implies the scaling relation
\begin{equation}
\alpha_s=\alpha+{\kappa_s}/z\,.
\end{equation}
Inserting the known estimates from Ref.~\cite{Lubeck04} this scaling relation 
yields the estimates
\begin{equation}
\alpha_s\;=\; \left\{
\begin{array}{ll}
0.86(4) & \mbox{ for } d=1 \\ 
1.14(3) & \mbox{ for } d=2 \\ 
1.51(9) & \mbox{ for } d=3 \\ 
\end{array}
\right.\,.
\end{equation}
In one dimension our result is in very good agreement with the estimate 
$\alpha_s=0.85(2)$ reported in Ref.~\cite{BonachelaMunoz08}.

\section{Heuristic proof for ${\kappa_s}=1$ in one dimension}

The numerical result ${\kappa_s} = 1{.}00(2)$ suggests that ${\kappa_s}$ 
might be exactly equal to~$1$. This conjecture can be supported as follows. 
It is argued in~\cite{Vespignani98a}
that the critical behavior of the Manna model can be described effectively 
by two mutually coupled Langevin equations of the form\footnote{In our opinion these 
Langevin equations should be considered as an educated guess. Derivations 
from microscopic models are available~\cite{BonachelaMunoz08} but they involve 
various approximations which are hard to verify.}
\begin{eqnarray}
\partial_t \rho(\xvec,t) &=& -r \rho(\xvec,t) - b \rho^{2}(\xvec,t) + \nabla^2 \rho(\xvec,t)\nonumber \\
&& + \omega \rho(\xvec,t) \phi(\xvec,t) 
+ \sigma \sqrt{\rho(\xvec,t)} \, \eta(\xvec,t)\label{eq:Langevin1}\\
\partial_t \phi(\xvec,t) &=& D \nabla^2 \rho(\xvec,t)\mbox{.}\label{eq:Langevin2}
\end{eqnarray}
Here $\rho(\xvec,t)$ describes the coarse-grained density of active sites 
while $\phi(\xvec,t)$ represents the coarse-grained background density of 
all particles. 
As usual, $\eta(\xvec,t)$ denotes an uncorrelated white Gaussian noise. 
Note that Eq.~(\ref{eq:Langevin1}) differs from the Langevin equation for 
DP by an additional term $\omega \rho \phi$ which couples the two fields 
whereas Eq.~(\ref{eq:Langevin2}) describes diffusive reordering of the 
background field during toppling. Furthermore note that in a system with periodic boundary conditions the structure of Eq.~(\ref{eq:Langevin2}) ensures global conservation of the background field since $\partial_{t} \int \d^dx\phi(\xvec,t) 
= D \int\d^dx  \nabla^2 \rho(\xvec,t)= 0$. 

In the following we show that the structure of these Langevin equations 
implies ${\kappa_s}=1$ in one dimension. 
The idea is that the removal of active particles at the boundary and its 
redistribution in the (ideally infinite) bulk of the system depletes not 
only in the coarse-grained activity $\rho(x,t)$ but also the 
background density $\phi(x,t)$. 
This is in fact the essential feature that makes boundary phenomena 
in the Manna class different from those of DP. 
Using the Langevin equation we compute the integrated loss of the 
background density and show that the results in one dimension would be inconsistent unless ${\kappa_s}=1$.

To this end we choose a particular site at distance $x_0$ and monitor the 
loss of the background density at this point. 
Since the depletion zone grows as $t^{1/z}$, 
this density will begin to decrease at about a typical time $t_0$ 
which scales as $x_0^z$. 
How much of the background density will be lost between $t=t_0$ and $t=\infty$? 
According to Eq.~(\ref{eq:Langevin2}), the total loss $\Delta\phi=\phi(x_0,t_0)-\phi(x_0,\infty)$ is given by the integral
\begin{equation}
	\Delta\phi\;=\; 
	-\int^{\infty}_{t_0} \partial_t \phi(x_0,t) \,\d t \;\stackrel{\mathrm{(\ref{eq:Langevin2})}} =\; 
	-\int^{\infty}_{t_0} D \,\partial^{2}_{x} \rho(x_0,t) \,\d t 
	\label{eq:GammaBeweis1}\,.
\end{equation}
Inserting the scaling form~(\ref{ScalingForm}) and substituting the scale-invariant ratio 
\begin{equation}
\xi = \frac{x_0}{t^{1/z}}
\end{equation}
we obtain
\begin{eqnarray}
	\Delta\phi &=& 
	-D\int^{\infty}_{t_0} t^{-2/z-\alpha}R''(x_0/t^{1/z}) \,\d t  \nonumber \\
	&=& -D z \,x_0^{z-z\alpha-2}\int_{0}^{\xi_0} \xi^{1-z+z\alpha} \, R''(\xi)\,\d \xi 
	\label{eq:GammaBeweis2}\,,
\end{eqnarray}
where $\xi_0 = x_0/t_0^{1/z}$. Note that this expression is positive since $R(\xi)$ has a negative curvature.

Since the total loss $\Delta \phi$ is a finite quantity, the integral~(\ref{eq:GammaBeweis2}) has to be finite as well. Obviously, the integral is finite if and only if the integrand diverges slower than $1/\xi$ as $\xi\to 0$. In this limit the scaling function can be approximated by $R(\xi)\sim\xi^{\kappa_s}$ so that the integrand diverges as
\begin{equation}
\xi^{1-z+z\alpha}\,R''(\xi) \;\simeq \; \kappa_s (\kappa_s-1)\,\xi^{-1-z+z\alpha+\kappa_s}\,.
\end{equation}
For this reason the integral is finite if one of the following three conditions holds:
\begin{quote}
\begin{itemize}
\item ${\kappa_s}=0$ or
\item ${\kappa_s}=1$ or
\item $\alpha_s=\alpha+\kappa_s/z > 1$.
\end{itemize}
\end{quote}
The first solution $\kappa_s=0$ is unphysical because there would be no depletion zone, meaning that we are left with the other two possibilities. Which of them applies depends on the dimensionality of the system: In one dimension the exponent $\alpha_s = 0.86(4)$ is clearly smaller than~$1$, hence the above integral is finite if and only if ${\kappa_s}=1$. In higher dimensions, however, $\alpha_s$ is larger than $1$ so that $\kappa_s$ 
can take arbitrary values. This explains why $\kappa_s=1$ in one-dimensional systems only, reflecting particle conservation.

\section{Mean field approximation}

In $d > 4$ dimensions, i.e. above the so-called upper critical dimension $d_c=4$, 
the Manna model at criticality is expected to display mean field behavior on 
large scales which is usually characterized by simple values of the exponents. 
In the following we compute the surface exponent $\alpha_s$ in the mean field limit, 
verify the validity of the scaling form~(\ref{ScalingForm}), 
and show how the corresponding partial differential equations can be solved numerically.

\begin{figure}[t]
\centering\includegraphics[width=150mm]{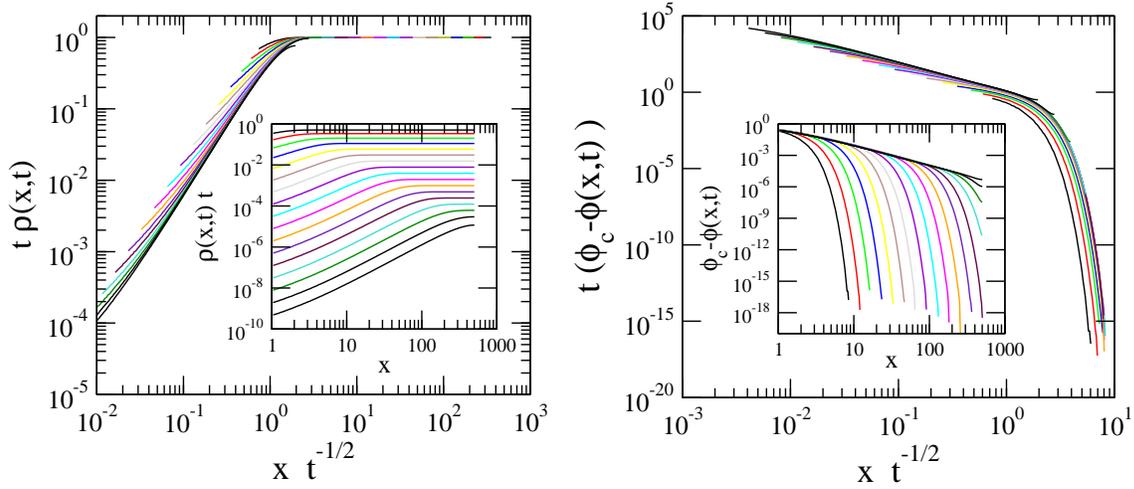}
\caption[Numerical integration of the discretized mean-field equations]{Numerical integration of the mean field equations~(\ref{mfeq1}) and~(\ref{mfeq2}) at the critical point. The left and the right panel show results for the density of active sites $\rho(x,t)$ and the deviation of the background density of particles $\phi(x,t)$ from its critical values $\phi_c$ measured at $t=1,2,4,8,\ldots,65536$. The data is rescaled according to the expected scaling forms while the insets display the raw data.}
\label{fig:mf-coupled}
\end{figure}

As usual, we assume that the mean-field equations have the same form as the 
Langevin equations~(\ref{eq:Langevin1})-(\ref{eq:Langevin2}) if the noise 
term is neglected. 
Since the boundary breaks translational invariance, space-dependence has to 
be retained. Moreover, the time scale 
of the process may be fixed by choosing $b=1$. 
Furthermore, the parameter $D$ can be absorbed in $\omega$ via rescaling of $\phi$ 
and thus we may set $D=1$. 
The resulting set of mean field equations reads
\begin{eqnarray}
	\partial_t \rho(\xvec,t) &=& -r \rho(\xvec,t) - \rho^{2}(\xvec,t) + \nabla^2 \rho(\xvec,t) + \omega \rho(\xvec,t) \phi(\xvec,t)\mbox{,}\label{eq:LangevinStart1}\\
	\partial_t \phi(\xvec,t) &=& \nabla^2 \rho(\xvec,t) \mbox{.}\label{eq:LangevinStart2}
\end{eqnarray}
If the boundary has the form of a $d-1$-dimensional hyperplane it is clear 
that the mean field densities will only depend on the distance perpendicular 
to the wall. Denoting this distance by $x$ we may replace the gradient by an ordinary 
derivative perpendicular to the wall, leading to
\begin{eqnarray}
\label{mfeq1}
\partial_t \rho &=& - \rho^{2} + \partial^2_x \rho + (\omega \phi-r)\rho \mbox{,}\label{dgl1}\\
\label{mfeq2}
\partial_t \phi &=& \partial_x^2 \rho\label{dgl2} \,,
\end{eqnarray}
where we omitted the arguments $x,t$. Obviously the critical background field is $\phi_c=r/\omega$.
We iterated these partial differential equations numerically by means of standard methods (see Fig.~\ref{fig:mf-coupled}). Because of the scaling form~(\ref{ScalingForm}) 
\begin{eqnarray}
\label{SForm1}
\rho(x,t) \;\simeq\; t^{-\alpha} R\left(\xi\right)
\end{eqnarray}
with the scale-invariant variable $\xi=x/t^{1/z}$ one expects a data collapse if $t^\alpha\rho(x,t)$ is plotted against $\xi$. As shown in the left panel of Fig.~\ref{fig:mf-coupled}, a reasonable collaps is obtained for $\alpha=1$ and $z=2$, the well-known bulk exponents of DP in the mean field limit. Similarly, the background density should obey a scaling form
\begin{equation}
\phi(x,t) \;\simeq\; \phi_c-t^{-\alpha'} F\left(\xi\right)
\end{equation}
where the numerical integration (see right panel of Fig.~\ref{fig:mf-coupled}) leads to the value $\alpha'=1$. This is reasonable since the first and the third term in Eq.~(\ref{mfeq1}) should be equally relevant, hence $\alpha=\alpha'$ in the mean field limit.

Although the data collapses become gradually better for large $t$, they are not really perfect, raising the question whether slightly different critical exponents would lead to a better data collapse. However, this would be not consistent with the structure of the mean field equations~(\ref{mfeq1}) and~(\ref{mfeq2}). To see this we eliminate the parameter $r$ by a shift of $\phi\to\phi+r/\omega$, solve the first differential equation for $\phi$ and insert it into the second one, arriving at a single partial differential equation for the density of active sites $\rho$
\begin{equation}
(\partial_t \rho) (\partial_x^2\rho - \partial_t \rho )  \;=\;
\rho^2 (\omega\partial_x^2\rho-\partial_t\rho)  + \rho(\partial_x^2\partial_t\rho-\partial_t^2\rho) \,.
\end{equation}
Inserting the scaling form~(\ref{SForm1}) one obtains an autonomous 
ordinary differential equation for $R(\xi)$ 
(meaning that it does no longer depend explicitely on $x$ and~$t$) 
if and only if $\alpha=1$ and $z=2$. This leads to a non-linear third-order differential equation of the form
\begin{eqnarray}
\label{eq:autonomie}
 R'''(\xi ) &=& \frac{1}{2\xi } \, \Bigl[
2 R(\xi ) \Bigl(2 \omega  R''(\xi )+\xi  R'(\xi
   )-2\Bigr)\\
&& \hspace{7mm} +\xi  \frac{R'(\xi )}{R(\xi)} \Bigl(2 R''(\xi )+\xi  R'(\xi\nonumber
   )\Bigr)\\
&& \hspace{7mm} -\left(\xi
   ^2+4\right) R''(\xi )+3 \xi  R'(\xi )+4 R^2			(\xi )\nonumber 
\Bigr]\,.
\end{eqnarray}
The trivial solution $R=0$ is unphysical while the solution $R=1$ describes the homogeneous case without boundary. 
With an absorbing  wall one has to impose a Dirichlet boundary condition $R(0)=0$ 
together with the bulk limit $R(\infty)=1$. 
The second term $\frac{R'(0)R''(0)}{R(0)}$ in the second line of Eq.~(\ref{eq:autonomie}) requires that either $R'(0)$ or $R''(0)$ vanishes, hence the solution $R(\xi)$ is either linear or quadratic at the origin. The linear solution can be excluded because it would imply that the loss of the background density according to the second Langevin equation~(\ref{eq:Langevin2}) would be zero. Therefore, the solution behaves as $R(\xi) \propto \xi^2$ for small $\xi$, hence we obtain the exponent $\kappa_s=2$. 

We did not succeed in solving Eq.~(\ref{eq:autonomie}) analytically. However, Eq.~(\ref{eq:autonomie}) can be iterated numerically by setting $R(0)=R'(0)=0$ and tuning $R''(0)\approx 0.283272$ in such a way that the iteration approaches $R(\xi)=1$ for large $\xi$. The data (not shown here) is in perfect agreement with the directly iterated mean field equations.

\section{Conclusions}

In the present paper we have investigated the properties of an absorbing hyperplane in a critical Manna model. All results are consistent with earlier findings by Bonachela \etal~\cite{BonachelaMunoz07,BonachelaMunoz08} in 1d, who used such boundary effects as a criterion to discriminate between directed-percolation and Manna scaling. We have extended these numerical studies to higher dimensions and have confirmed that the observed scaling properties are consistent with the mean field limit. Moreover we have shown that in one spatial dimension, the structure of the Langevin equations, which are believed to describe Manna-type critical behavior on a coarse-grained scale, allows one to calculate the surface-critical exponents by means of a scaling relation.


\vspace{5mm}
\noindent\begin{large}\textbf{References}\end{large}

\bibliographystyle{revtex}
\bibliography{master}

\end{document}